\documentclass[12pt]{article}
\pdfoutput=1  
\newcommand{\TITLE}{Elastic hadron scattering and optical theorem}
\newcommand{\KEYWORDS}{optical theorem, short-ranged hadron interaction, elastic scattering}

\usepackage{amssymb,amsmath,latexsym}
\usepackage[a4paper, twoside,
            top=2.5cm,bottom=2.5cm,
            left=2.4cm,right=2.4cm,
            bindingoffset=1.0cm
            ]{geometry}

\usepackage{hyperref}
\hypersetup{
    pdftex,
    bookmarks=true,         
    bookmarksnumbered=true,
    bookmarksopen=true,
    bookmarksopenlevel=0,
    bookmarksnumbered=true,
    pdftitle={\TITLE},    
    pdfauthor={M. V. Lokajicek, V. Kundrat, J. Prochazka},      
    pdfsubject={HEP-theory}, 
    pdfkeywords={\KEYWORDS}, 
}

\usepackage{cleveref}  
\newcommand{\refp}[1]{(\ref{#1})}

\crefname{equation}{Eq.}{Eqs.}
\crefname{section}{Sec.}{Secs.}
\crefname{chapter}{Chapter}{Chapters}
\crefname{table}{Table}{Tables}
\crefname{figure}{Fig.}{Figs.}
\crefname{appsec}{Appendix}{Appendices}
\crefname{appchap}{Appendix}{Appendices}

\newcommand{\ket}[1]{\left| #1 \right>} 
\newcommand{\bra}[1]{\left< #1 \right|} 

\renewcommand\Re{\operatorname{Re}}
\renewcommand\Im{\operatorname{Im}}
\makeatletter
\def\blfootnote{\xdef\@thefnmark{}\@footnotetext}
\makeatother

\begin{document}
\begin{center}
\blfootnote{
{\hspace{-8mm}\it * Corresponding author\\}
{\it Email addresses:} lokaj@fzu.cz (Milo\v{s} V.~Lokaj\'{\i}\v{c}ek), kundrat@fzu.cz (Vojt\v{e}ch Kundr\'{a}t), jiri.prochazka@cern.ch (Ji\v{r}\'{\i} Proch\'{a}zka)} 
{\Large\bf \TITLE} \\[8mm]

Milo\v{s} V.~Lokaj\'{\i}\v{c}ek$^{\text{a}}$, Vojt\v{e}ch Kundr\'{a}t$^{\text{a}}$ and Ji\v{r}\'{\i} Proch\'{a}zka$^{\text{b,}}$\footnote{On leave of absence from Institute of Physics, AS CR, v.v.i., 182 21 Prague 8, Czech Republic}$^{\text{,*}}$
\\[4mm]
{\it   
   $^{\text{a}}$Institute of Physics of the AS CR, v.v.i., 18221 Prague 8, Czech Republic\\

   $^{\text{b}}$CERN, Geneva, Switzerland} \\
\vspace{4mm}
\end{center}
\vspace{6mm}

\noindent
{\bf Abstract}\\
All contemporary phenomenological models of elastic hadronic scattering have been based in principle on the assumption of optical theorem validity that has been overtaken from optics. It will be shown that the given theorem which has not been actually proved in particle physics cannot be applied to short-ranged strong interactions. The analysis of corresponding collision experiments is to be done under new basic assumptions. The actual progress in description of hadronic collision processes might then exist only if the initial states are specified on the basis of impact parameter values of colliding particles and probability dependence on this parameter is established, without limiting corresponding conclusion by optical theorem validity. \\

\noindent
{\bf keywords:} \KEYWORDS

\section{\label{sec:introduction}Introduction}
Practically all hitherto models of elastic hadronic scattering have been based on the assumption of optical theorem validity. According to this theorem the total (hadronic) cross section $\sigma_{\text{tot}}$ is proportional to the imaginary part of elastic (hadronic) scattering amplitude $f$ at zero scattering angle $\theta$
\begin{equation}
\sigma_{\text{tot}} \propto \Im f(\theta=0)
\label{eq:optical_theorem_prop}
\end{equation}
when the complex amplitude $f(\theta)$ is obtained, e.g., with the help of Schroedinger equation. 

The given theorem used now in particle physics was overtaken from the optics where it developed from the formula for refraction index (defined on the basis of wave theory of light) which contained also the influence of extinction cross section (now denoted as total cross section); see the story described by Newton \cite{newton1975}. The formula~\refp{eq:optical_theorem_prop} has been formulated practically only on the basis of experimental refraction data without theoretical reasoning.

Different attempts to prove it theoretically have been done mainly when the collisions of fundamental particles have started to be studied and the optical theorem has been applied to also in the region of strong interactions. Some of these attempts have been interpreted as successful. However, we have demonstrated recently that fundamental discrepancy has been to exist especially if the optical theorem has been applied to elastic scattering caused by the very short-ranged strong interaction \cite{Lokajicek_optical_theorem}.

Some arguments used to support its validity in strong interactions have been, however, still repeated. In the following we shall attempt to provide some deeper and more systematical reasoning why this theorem can be hardly applied to in any hadronic elastic scattering. We shall demonstrate it on two arguments that have been described, e.g., in \cite{barone2002}: one based on the optical approach (wave interpretation) and the other on unitarity of corresponding $S$ matrix. However, in both the cases the corresponding conclusions have been based evidently on some additional assumptions that cannot be applied to strong hadronic collisions in any case. 

The argument based on $S$ matrix theory has been nearer to the theory of strong interactions. However, the corresponding conclusions have been based on the assumptions concerning the basic structure of $S$ operator acting in the Hilbert space in which the incoming and outgoing states cannot be correspondingly distinguished. We shall start, therefore, by discussing the necessary Hilbert structure formed by Schroedinger equation solutions in corresponding collision processes; see \cref{sec:schroedinger_hilbert}. The two mentioned approaches trying to prove the validity of optical theorem will be then analyzed in \cref{sec:sch_eq_optical_theorem}; the assumptions being in contradiction to collision characteristics for strong interaction will be specified.
   
As to the contemporary models of elastic collisions they have corresponded to quite phenomenological mathematical description of corresponding collision processes. More detailed physical description may be obtained if the statistical distribution of impact parameter values between two colliding objects is taken into account and the dependence of collision characteristics on this parameter is established. Basic aspects of corresponding probabilistic model proposed recently will be briefly described in \cref{sec:impact_parameter_description}. The model clearly shows that the collision process may be interpreted on fully ontological basis (without applying optical theorem). 

\section{\label{sec:schroedinger_hilbert}Schroedinger equation and corresponding Hilbert space}
Time evolution of microscopic processes is being described with the help of the Schroedinger (linear differential) equation
\begin{equation}
i\hbar\frac{\partial\psi(x,t)}{\partial t} = \hat{H} \psi(x,t)
\label{eq:schr_equ}
\end{equation}
where Hamiltonian operator $\hat{H}$ is given by
\begin{equation}
\hat{H} = -\frac{\hbar^2}{2m}\nabla^2 + \hat{V}(x)
\end{equation}
where $\hat{V}(x)$ is corresponding potential.
Its basic solutions (represented by the product of space and time functions) may be expressed in the form
\begin{equation}
      \psi_E(x,t) = \lambda_E(x)e^{-iEt/\hbar};
\label{eq:sch_equ_partial_solution}
\end{equation}
being standardly normed to one:  $\int \text{d}x\, \psi^*_E(x,t)\psi_E(x,t)=\int \text{d}x\, \left| \lambda_E(x)\right|^2=1$ (at any time $t$). The function $\lambda_E(x)$ of all space coordinates ($x$) corresponds to the Hamiltonian eigenfunctions
\begin{equation}
 \hat{H} \lambda_E(x) = E \lambda_E(x) \label{eq:schr_eq_time_independent}.
\end{equation}
General solution $\psi(x,t)$ of Schroedinger equation~(\ref{eq:schr_equ}) may be then written as a superposition of corresponding Hamiltonian eigenfunctions $\lambda_E(x)$
\begin{equation}
\psi(x,t) =  \sum_E c_E  \psi_E(x,t)
\end{equation}
where $c_E$ are corresponding coefficients in a linear combination of particular solutions $\psi_E(x,t)$; fulfilling  $\sum_E |c_E|^2=1$.

All possible amplitudes $\psi(x,t)$ (functions of space coordinates at different $t$ values) form then a complete Hilbert space. Schroedinger defined then expected values $A(t)$ of physical quantities
\begin{equation}
   A(t) = \int\psi^*(x,t)\,\hat{A}\,\psi(x,t)dx    
\label{eq:expected_value}
\end{equation}
corresponding to classical quantities. It was shown originally for inertial motion only; however, it holds practically generally (see \cite{inte}). Only the set of Schroedinger solutions is smaller due to discrete quantum states in closed systems. It was shown by Ioannidou \cite{ioannidou1982} and Hoyer \cite{hoyer2002} that the Schroedinger equation might be derived from statistical combination of Hamilton equation solutions (or be at least equivalent to these solutions) if their whole set was limited by a rather weak condition; see also \cite{inte,adv}.

A $t$-dependent solution $\psi_E(x,t)$ of Schroedinger equation (the set of vectors in the corresponding Hilbert space corresponding to different values of $t$) represents the evolution of motion as an open trajectory in the case of continuous energy spectrum or as a closed trajectory for discrete energy values. Each physical quantity $A(t)$ is then represented by associated operator $\hat{A}$ acting in the given Hilbert space. 

Any vector $\psi_E(x,t)$ represents then instantaneous state belonging to two opposite momentum directions. To distinguish these two different cases the total Hilbert space (in the case of elastic collisions) must consist of two mutually orthogonal subspaces each being spanned on the basis of Hamiltonian eigenfunctions $\lambda_E(x)$ as it has been shown already many years ago by Lax and Phillips \cite{lax1,lax2} and independently derived also by Alda et al.~\cite{alda} from the requirement of exponential (purely probabilistic) decay law of unstable particles. Only in such an extended Hilbert space the collision processes of two particles may be correspondingly described. 
 The transition from one subspace to another may be then given by the evolution operator 
\begin{equation}
     \hat{U}_{ev}(t) \; =\; e^{-i\hat{H}t/\hbar}; 
\label{eq:ev_op}
\end{equation}
 the opposite evolution corresponding to negative values of $t$. It holds then
\begin{equation}
     \psi_E(x,t) \;=\; \hat{U}_{ev}(t)\psi_E(x,0) \,.
 \end{equation}
If $\psi_E(x,0)$  represents the state corresponding to the shortest distance between two colliding particles then the states for time $t>0$ belong  to the subspace of outgoing particles and for time $t<0$ to that of incoming states.
 
The given Hilbert structure has been, however, excluded by Bohr in 1927; he asked for the Hilbert space of any physical system to be spanned always on one basis of Hamiltonian eigenvectors. It has caused that the earlier physical interpretation of Schroedinger equation solutions has been fundamentally deformed as any description of continuous time evolution has been practically excluded. Moreover, the given model has required the existence of immediate interaction between very distant particles, which was shown and criticized by Einstein in 1935 with the help of special coincidence Gedankenexperiment. The physical scientific community preferred, however, Bohr's approach (in the region of microscopic processes). 
    
Later both the alternatives were admitted and discussed. Bohr's alternative was, however, supported again on the basis of the fact that Bell's inequality (derived in 1964 for the coincidence experiment more specified than that of Einstein) was violated in the corresponding experiment including spin measurement and performed by Aspect et al.~in 1982 \cite{aspect1982}. It has been shown only recently that Bell's inequality was based always on an assumption that did not hold in the given more specified experiment (but only in that proposed originally by Einstein); see, e.g., \cite{inte}. Consequently, Einstein has been fully right in the given controversy with Bohr and the Hilbert space must always consist at least of two mutually orthogonal subspaces as explained in the preceding. All necessary details may be found in \cite{inte,scripta,Lokajicek2013_intech} and \cite{Lokajicek2013_bell}.

\section{\label{sec:sch_eq_optical_theorem}Two attempts of deriving optical theorem }
In the  region of strong interactions the decisive study of elastic processes concerns two proton collisions where the experimental data especially for small scattering angles represent the combination of Coulomb and hadronic interactions. The ratio of these two interactions has always being  determined on the basis of some theoretical predictions. However, the contemporary approaches (in both the interaction kinds) have started often from some assumptions that do not correspond to physical reality as it will be shown in the following. 

As to the Coulomb interaction it has been assumed that the corresponding elastic differential cross section has risen to infinity for very small scattering angles, which has followed from the fact that the zero scattering angle should be obtained at infinite distance (i.e., at infinite impact parameter). However, the measured region of scattering angles corresponds to impact parameters of less than micrometers, which is not respected in the usual formula that is used for interpretation of Coulomb part of measured data. In addition to, a part of measurable elastic collisions may be caused by multiple Coulomb scattering according to experimental conditions (target density).

The similar criticism concerns, of course, the assumed behavior of strong interactions in the same region. Here, the validity of optical theorem given by \cref{eq:optical_theorem_prop} has been assumed practically in all theoretical as well as experimental studies. The optical theorem has been overtaken from optics without having been proved in the past. It will be shown that also all contemporary attempts to prove its validity in the case of strong interactions have been based on assumptions that are not surely valid in the case of strong interaction. 

As it has been already mentioned there are two main approaches that have been used in attempts to derive the optical theorem for elastic scattering of two particles. The main attempt to derive it has been done in the framework of $S$ matrix theory when some important assumptions have concerned the structure of corresponding Hilbert space as well as of $S$ matrix itself (see, e.g., \cite{barone2002}). In the other approach (introduced also in \cite{barone2002}) the ambition to derive the given theorem has been based  on the wave theory. Both the approaches will be analyzed in two following subsections. In the third subsection the consequences of different mechanisms of electromagnetic and strong interactions for the solution of the given problem will be mentioned and discussed.  

\subsection{\label{sec:s_matrix_transition}S matrix theory and transition operator} 

The S-matrix theory tried to describe a collision process with the help of $S$ operator acting in a Hilbert space spanned on all possible states and defined in principle in a phenomenological way. The $S$ operator has been then assumed to transform initial state $\ket{i}$ directly to final one $\ket{f}$: 
\begin{equation}
    \ket{f} = S \ket{i} \,.
\end{equation}
The probability of corresponding transition has been given by matrix element
\begin{equation}
   P_{i\rightarrow f} = \left|\bra{f}S\ket{i}\right|^2.    
\label{eq:p_if}
\end{equation}
The given $S$ operator has been then required to fulfill the condition of unitarity
\begin{equation}
     S^+S = S\,S^+=\,I \;.
\label{eq:s_unitary}
\end{equation}

Practically in all approaches attempting to derive optical theorem the $S$ matrix has been defined in the form (see, e.g., \cite{barone2002}, p.~52)
\begin{equation}     
     S = I + iT   
\label{eq:s_matrix}
\end{equation}
where the introduction of unit matrix $I$ has been based necessarily on the assumption that final collision events have always been represented by the same state set as initial states. It has then followed from the unitarity of this $S$ operator (\cref{eq:s_unitary}) and \cref{eq:s_matrix} 
\begin{equation}
   T^{+}T \,= \,i\,(T^{+} - T).
\label{eq:T_condition}
\end{equation} 

From this equation the usual optical theorem given by \cref{eq:optical_theorem_prop} has been derived under further additional assumptions; e.g., the final and initial states have been taken as identical.  
There is, however, problem with the definition of initial states being identical to the final states if the different deviations from original direction are measured in elastic short-ranged collisions while only one (singular value) of these values, i.e., $\theta=0$, is to be attributed to the whole set of initial states. Any other initial states cannot exist under usual conditions. 

More detailed analysis of \cref{eq:p_if,eq:s_unitary,eq:s_matrix} allows us to derive then following conditions for corresponding probabilities
\begin{equation}
   \sum_f P_{i\rightarrow f} \;=\; 1 - 2 \Im T_{ii} + \sum_f |T_{if}|^2  \;=\; 1 
\end{equation} 
or
\begin{equation}
          \Im T_{ii}  \;=\; \frac{1}{2}\sum_f |T_{if}|^2 
\label{eq:Tii_condition}
\end{equation}
which should hold for any $i$; the last condition may be also derived from Eq.~(4.51) of \cite{barone2002} for a final state being identical to the initial one). The transition $i\!\rightarrow\!i$ is to be interpreted as an event when any collision process has not existed (or has been fully negligible). The condition~\refp{eq:Tii_condition} requires then for the number of corresponding events to increase when the number of collision processes rises, which is undoubtedly a contradictory condition requiring practically  $T=0$; the given definition of $S$ matrix being admissible in perturbation approaches only. Anyway, the transition matrix added to unit matrix leads to quite unacceptable physical characteristics when the collision processes do not represent only a small perturbation.  
It means that in the case of strong interactions the $S$ matrix cannot be defined generally by the condition~\refp{eq:s_matrix} and \cref{eq:T_condition} has not any sense in this case.

It may be seen from \cref{sec:schroedinger_hilbert} that the $S$ operator should be defined rather as acting in Hilbert space consisting of two mutually orthogonal Hilbert subspaces, one containing different initial states and the other containing corresponding final ones: $H\,=\,H_i\oplus H_f$. 
However, if in addition to elastic processes also some other inelastic processes  exist the subspace of final states is to be divided into two orthogonal subspaces; the total Hilbert space being defined as
\begin{equation}
              H\,=\,H_i\oplus H_f^{\text{el}}\oplus H_f^{\text{inel}}\,
\end{equation}
where the incoming or outgoing states are represented always by one vector.

The $S$ operator should then define the transition probabilities from an initial state to some states belonging to one of two divers final subspaces; no opposite transitions existing. The subspace $H_f^{\text{el}}$ represents the states of the same particle pair as $H_i$; however, they are to be characterized differently: incoming states are to be characterized by different impact parameter values while outgoing ones by scattering angle or equivalently. 
If we start from the ontological interpretation of collision processes (see the end of \cref{sec:schroedinger_hilbert}) it is necessary to expect that the Coulomb and strong interaction will behave very differently. While all events represented by vectors in the subspace $H_f^{\text{el}}$ will contribute to integrated elastic cross section in the former case the situation in the case of strong interaction will be quite different.

Hadronic collisions at a given impact parameter value may lead to an interval of scattering angles (according instantaneous space orientations of colliding protons). It is natural to assume in both the Coulomb and hadronic cases that the average scattering angle increases with decreasing impact parameter value.
However, for short-ranged strong interaction only a very small part of events (corresponding to impact parameters less than several femtometers) will contribute to total hadronic cross section while a much greater part of events (corresponding to impact parameters of greater values) will pass without any interaction. In the standard models (involving optical theorem validity) this part is, however, always added to elastic hadronic collisions.

As to the inelastic processes they are represented by transitions to $H_f^{\text{inel}}$. These final states may be interpreted  by one vector, too. In principle at least two different objects are to come into being, from which at least one is to be unstable,
decaying then into other objects.
The transition probabilities from $H_i$ to  $H_f^{\text{el}}$ or $H_f^{\text{inel}}$ are then defined by non-zero elements of unitary $S$ matrix.   

The existence of mutually orthogonal Hilbert subspaces has been considered also by Kupczynski \cite{Kupczynski2013}. However, there is a great difference between both these approaches. In our approach incoming and outgoing states belong to mutually orthogonal subspaces as it is required if incoming and outgoing particles are to be represented always by different vectors of corresponding Hilbert space (see \cref{sec:schroedinger_hilbert}) while in the system assumed in \cite{Kupczynski2013} each subspace should be divided into two further subspaces in such a case, yet. Moreover, in the corresponding Hilbert subspaces 
involving the production of more-particle final states the corresponding initial states are included, too, which are unphysical and which makes especially the unitarity condition of complete S matrix practically impossible.

In any case one can conclude from the preceding that the optical theorem is quite unacceptable for short-ranged strong interaction,
as quite arbitrary  values of total cross section may be obtained.
This arbitrariness may be much greater if at higher energy values the inelastic processes exist, too. Consequently, some new descriptions should be looked for in this case.

\subsection{\label{sec:s_matrix_optical_theorem}Derivation based on wave approach }

Other attempt to derive the optical theorem in strong interactions has been based on repeating the approach used in optics where a wave has been scattered by an obstacle. The collision process has been described with the help of wave amplitudes (see, e.g., p.~16 in \cite{barone2002}).  We shall not repeat the detailed approach here; only main points will be mentioned. 

The initial collision state has been represented by a plane wave  ($U_0$ is a constant)
\begin{equation}
    U_{\text{in}}(x,y,z) \, =\, U_0e^{ikz}. 
\end{equation}
The final state has been then expressed with the help of Fraunhofer diffraction, Babinet's principle and Huygens-Fresnel principle in the form of sum of unscattered and scattered events 
\begin{equation}
        U_f(x,y,z) \,=\, U_{\text{unsc}}(x,y,z) + \,  U_{\text{scatt}}(x,y,z) 
\label{eq:Af}
\end{equation}
The scattered wave has been expressed as          
\begin{equation}
        U_{\text{scatt}} \,=\,  U_0 f(q)\frac{e^{ikz}}{r}      
\label{eq:U_scatt}
\end{equation}
where $\vec{q}=\vec{k}'-\vec{k}$ is momentum transfer and $\vec{r}=(x,y,z)$ is a position vector ($|\vec{k}'|=|\vec{k}|=k$ in the case of elastic scattering); the squared modulus of $U_f(x,y,z)$ represents corresponding probability of outgoing wave, $U_{\text{unsc}}$ represents  the non-interacting part of original beam. The outgoing scattered states $U_{\text{scatt}}$ are characterized by vectors $\vec{q}$. 

It has been then written for the \emph{scattering amplitude}
\begin{equation}
     f(\vec{q}) = \frac{ik}{2\pi}\int\!\! \text{d}^2\vec{b}\;\Gamma(\vec{b})\;
                                 e^{-i\,(\vec{q}.\vec{b})}   
\label{eq:fq}
\end{equation}
where $\Gamma(\vec{b})$ has represented the profile of scattering center (obstacle). The intensity of the incident and of the scattered light has been taken as
\begin{equation}
I_{\text{in}} = |U_{\text{in}}|^2 = |U_0|^2
\end{equation}
and 
\begin{equation}
I_{\text{scatt}} = |U_{\text{scatt}}|^2 = |U_0|^2\frac{|f(\vec{q})|^2}{r^2}
\end{equation}
respectively. These definitions of intensities have been then used to define the elastic differential cross section as  
\begin{equation}
 \frac{\text{d}\sigma}{\text{d}\Omega} = \frac{I_{\text{scatt}}r^2}{I_{\text{in}}}=|f(\vec{q})|^2
\label{eq:dcs_wave}
\end{equation}

The integrated elastic cross section has been then equal (in approximation of small scattered angles $\theta\cong\sin\theta$)        
\begin{equation}
    \sigma_{\text{el}}=\frac{1}{k^2}\int|f(\vec{q})|^2\text{d}^2\vec{q}
    \cong  \int\!\! \text{d}^2\vec{b}\,|\Gamma(\vec{b})|^2 \,  .  \label{eq:sigma_el_waves}   
\end{equation}
Now it has been put $\; \Gamma(\vec{b})\,=\,1-S(\vec{b})\;$  and the last equation has been rewritten as (see Eq.~(2.37) in \cite{barone2002})
\begin{equation}
    \sigma_{\text{el}}=\int\!\! \text{d}^2\vec{b}\,|1-S(\vec{b})|^2 \,    \label{eq:S_el_waves}   
\end{equation}
on the basis of the assumption that the initial wave has equaled the sum of the passage through a hole and the diffraction caused by the object of the same shape. Then it has been possible to write for the absorption (inelastic) cross section
\begin{equation}
   \sigma_{\text{abs}}=  \int\!\! \text{d}^2\vec{b}[2\Re\Gamma(\vec{b})-|\Gamma(\vec{b})|^2]  \,. 
\label{eq:sigma_abs}
\end{equation}

In such a case it has been possible to express the total cross section in the form
\begin{equation}
     \sigma_{\text{tot}} = \sigma_{\text{el}}+\sigma_{\text{abs}} =
            2\int\!\! \text{d}^2\vec{b}\,\Re\Gamma(\vec{b}) 
\label{eq:sitot}
\end{equation}
and consequently (combining \cref{eq:sitot} with \cref{eq:fq} for $\theta=0$)
\begin{equation} 
     \sigma_{\text{tot}} = \frac{4\pi}{k}\Im f(\vec{q}=0) 
\label{eq:sgt}
\end{equation}  
which has been denoted as the optical theorem (see also \cref{eq:optical_theorem_prop}).

It is evident that in this wave approach all initial states are equally involved in the collision process. It may be hardly applied to strong interaction  
where a significant part of events goes mostly without any interaction. Also the existence of inelastic processes may be hardly included in the corresponding approach. It may be applied practically to Coulomb interaction only.
In such a case one may admit that a kind of optical theorem may be derived 
 for the elastic amplitude established with the help of corresponding Schroedinger equation; 
  especially, if eventually some acceptable limiting condition would be added, yet.   
           
\subsection{\label{sec:elmag_vs_strong}Difference of electromagnetic and strong interaction mechanisms }
    In the pp collisions the determination of elastic differential cross section from the measured data may be significantly influenced by one factor more. There is an important difference in the mechanism of Coulomb and strong interactions. While the Coulomb interaction acts at greater distances between colliding protons and may be described with the help of corresponding potential the strong interaction should be interpreted rather as contact one and the corresponding description of its effect should be looked for. 
 
 Consequently, the probability of strong collision at actual (initial) impact parameter value may be influenced significantly by the continuous effect of Coulomb interaction at greater distances before a proper (contact) collision may happen. The given influence may be rather different at divers collision energy values, the frequency of strong interaction events corresponding to initial impact parameter being decreased or increased according to charges of colliding particles (differently in dependence on collision energy).
  
There is not any doubt that the probability of a given hadronic collision process may be influenced fundamentally by corresponding impact parameter value if both the kinds if interactions are involved. It is then fully entitled to assume that incoming colliding particles are equally distributed in cross plane, the frequency of initial impact parameters increasing lineary with their values. The actual minimum mutual distance (in collision instant) of strongly colliding particles may be then significantly influenced by distant Coulomb interaction especially at lower collision energy values. It means that, e.g., for pp collisions the established value of total hadronic elastic cross section  may be lower than  it corresponds to reality. The analysis of earlier published results should be, therefore, examined also under these new conditions; see the next section 


\section{\label{sec:impact_parameter_description}Impact parameter description of collision processes} 

The hitherto models of elastic nucleon collisions have been in principle phenomenological, looking for the simple description of main scattering characteristics. However, when one is to understand better corresponding physical mechanism the distribution of at least some initial state characteristics must be taken into account. In collision processes the uniform statistical distribution of individual tracks around the common center-of-mass of colliding particles may represent important one. The frequency of processes corresponding to different impact parameter values should be taken into account in final collision formulas. 

Such a description trying to take into account realistic behaviour in the impact parameter space has been proposed by us in 1994 \cite{KLunpolarized1994}, see also \cite{Kundrat_2002,Kaspar_2011}. However, even if it has been possible to study some new characteristics of elastic collisions on the basis of impact parameter the deformation caused by assuming the commonly accepted optical theorem validity has remained until now. Its invalidity has been discovered fully only recently. 

If the limitation given by optical theorem is not applied to a quite new approach may be made use of in describing elastic collision processes. The corresponding collision model of two protons has been recently proposed by us in \cite{probmod}. Starting from the ontological interpretation of colliding objects and assuming that these objects are not fully spherical (differently oriented in space) one should expect that the probability of collision processes will depend mainly on the values of mutual impact parameter $b$ which should be uniformly distributed in corresponding cross plane. For the probability of elastic hadronic collisions at given impact parameter $b$ it may be then written:
\begin{equation}
   P^{\text{el}}(b) \;=\; P^{\text{tot}}(b)\,P^{\text{rat}}(b) 
\end{equation}
where $P^{\text{tot}}(b)$ is the probability of all possible hadronic (elastic or inelastic) collision processes and $P^{\text{rat}}(b)$ is the mutual ratio of elastic to total probabilities at corresponding value of impact parameter $b$. 

In the case of short-ranged (contact) strong interactions one can expect further on the basis of ontological realistic approach that elastic collisions will be mainly peripheral. The functions $P^{\text{tot}}(b)$ and $P^{\text{rat}}(b)$ may be then assumed to be monotonous functions of $b$: the first one diminishing with rising $b$ and the other increasing in the same interval of $b$.  Both the monotonous functions may be determined from corresponding experimental differential elastic cross section if one admits that a proton may exist at some internal states differing at least very slightly in their dimensions. 

The new collision model has been applied (in its preliminary form) to experimental data represented by measured elastic proton-proton differential cross section at energy of 52.8~GeV \cite{Lokajicek2013_intech}. It has been possible to demonstrate explicitly that some new possibilities of fundamental particle research have been opened on this basis; including also some preliminary new characteristics of proton in dependence on impact parameter, see \cite{Lokajicek2013_intech} for more details.

In this preliminary analysis the influence of distant Coulomb interaction has not been yet taken into account. The more detailed analysis of proton-proton collisions under new conditions is being prepared. 

\section{\label{sec:conclusion}Conclusion}
The optical theorem has been commonly applied to in description of elastic hadronic collision processes in the past even if all attempts to prove its validity in particle physics have been based on very limiting assumptions. These assumptions might be acceptable for long-ranged interaction like the Coulomb one but it may be concluded from the preceding that the given theorem is inapplicable in the case of short-ranged strong interaction. 

The whole problem is, of course,  rather complicated as the optical theorem concerns one point ($\theta=0$) of elastic differential cross section, lying in the interval of non-measurable deviations. The determination of the given limit is practically always burdened further by the fact that the influence of Coulomb effect that is to be subtracted is probably much greater than that of strong interaction. The commonly used parameterization (based fully on the assumption of optical theorem validity) of the function representing elastic differential cross section represents then
 important unphysical limitation, too.      

To respect the ontological characteristics of elastic collision processes the initial and final states are to be represented in two mutually orthogonal subspaces of the Hilbert space formed by the solutions of corresponding Schroedinger equation (see \cref{sec:schroedinger_hilbert}). Only then the influence of particle dimensions may be fully respected; the statistical distribution of impact parameter values for initial states and that of angle deviations for final states being represented in individual subspaces.
New elastic collision model \cite{Lokajicek2013_intech} based on these requirements has been shortly characterized in \cref{sec:impact_parameter_description}. It might open a deeper insight concerning the characteristics of hadronic collision processes and proper hadronic structure.

\end{document}